\documentstyle[twoside,fleqn,espcrc2,epsf]{article}

\newcommand{\AmS}{{\protect\the\textfont2
  A\kern-.1667em\lower.5ex\hbox{M}\kern-.125emS}}

\title{
\vskip -2.0cm
\hfill {\normalsize TTP94--09}\\
\vskip -0.2cm
\hfill {\normalsize June 1994}\\
\vskip 1.0cm
Top Quark Physics}

\author{M. Je\.zabek
        \address{Institut f\"ur Theoretische Teilchenphysik,
        Universit\"at Karlsruhe,
        D-76128 Karlsruhe, Germany}
        \address{Institute of Nuclear Physics,
        ul.Kawiory 26a, PL-30055 Cracow, Poland}%
        \thanks{Work partly supported by KBN under contract
            2P30225206 and by DFG under contract 436POL173193S.}}
\begin{document}

\begin{abstract}
Top quark studies at future $e^+e^-$ colliders
are considered.
Two issues are discussed:
{a -- }Some results are presented on the decays of top quarks.
Energy distributions of charged leptons and neutrinos
in  $t\to bW\to be^+\nu$  and jets in $t\to bW\to b\bar du$ decays
are sensitive to the structure of $tbW$ vertex.
Distributions of charged leptons from top decays are
particularly useful in polarization studies whereas neutrinos are
sensitive to deviations from the Standard Model.
{b -- }Recent calculations  are reviewed on the
top quark pair production in $e^+e^-$ annihilation.
The differential cross sections in the threshold region
can lead to an accurate determination of the top quark
mass and the interquark potential. The effects of the top-Higgs
Yukawa coupling and some higher order QCD corrections are
also under control.
\end{abstract}

\maketitle

\section{TOP QUARK AT $\bf e^+e^-$ COLLIDERS}

\thispagestyle{empty}
Top quark is the heaviest fermion of the Standard Model. The existing
direct lower mass limit\cite{D0} $m_t\ge 131$ GeV
as well as the value
$m_t = 174{}^{+11}_{-12}{}^{+17}_{-19}$ GeV
derived from the recent fit to LEP+SLC data \cite{Blondel} indicate
that its production is out of reach of LEP~II. (Two weeks after this
talk had been delivered the top quark saga culminated at the press
conference at FNAL when evidence for $t\bar t$ production was
announced by the CDF Collaboration \cite{CDF}.
The reported value of the
top quark mass $m_t = 174\pm{10}{}^{+17}_{-19}$ GeV
agrees very well with the indirect determination
\cite{Blondel}.) \par
\begin{figure}[htb]
\epsfxsize=7.5cm
\leavevmode
\epsffile[15 140 580 700]{mjfig1.ps}
\vskip-0.5cm
\caption{Total cross sections for the reaction
$e^+e^-\to e^-\bar\nu t\bar b(e^+\nu \bar t b)$
as the function of the top quark mass \protect\cite{Boos}.}
\label{fig-boos}
\end{figure}
It has been argued \cite{RV} that single top
production may be observed for $m_t$ up to about $180$ GeV if
LEP~II center-of-mass energy is pushed to $\sqrt{s}=210$ GeV.
However, a recent article \cite{Boos} found
a dramatic cancellation
of contributions to the reaction
$e^+e^-\to e^-\bar\nu t\bar b$.
A similar destructive interference
damps the single top production in $e\gamma$ collisions
\cite{Boos}. It is noteworthy, however, that this does not occur
for $p\bar p$ and $pp$ collisions \cite{WD} where for large
top masses the cross section of single top production
becomes comparable to that of $t\bar t$ production.
In Fig.\ref{fig-boos} the cross sction of reaction
$e^+e^-\to e^-\bar\nu t\bar b$
is shown for $\sqrt{s}=$170, 190 and 210 GeV \cite{Boos}.
The solid lines correspond to the complete set
of Feynman diagrams, the dashed lines to the photon exchange
subset and the dotted lines to the Weizs\"acker-Williams
approximation.
It is evident that even for the integrated luminosity
${\cal L}=500\ pb^{-1}$ and $\sqrt{s}=210$ GeV
one expects no events. Therefore, the top quark can be observed
in $e^+e^-$ collisions only if the
{\bf N}ext {\bf L}inear {\bf C}ollider is built.
Comprehensive studies have been performed of the physics
potential of a 500 GeV center-of-mass energy collider.
The results are summarized in  \cite{Saar,Hawaii,DESYWac}.
Theoretical studies on the top quark are reviewed in
\cite{ZerKue,ZerFin,ThA,KuHaw}.
\begin{table*}[hbt]
\setlength{\tabcolsep}{1.5pc}
\newlength{\digitwidth} \settowidth{\digitwidth}{\rm 0}
\catcode`?=\active \def?{\kern\digitwidth}
\caption{Top quark width as the function of $m_t$ including
($\Gamma_{t}$) and without ($\Gamma^{(0)}_{nw}$)
the contributions of the QCD, non-zero $W$ boson width (WBW)
and electroweak (EW) corrections.}
\label{tab:totwid}
\begin{tabular*}{\textwidth}{@{}@{\extracolsep{\fill}}c|c|cccc}
\hline
$\qquad m_t$[GeV]$\quad$ & $\Gamma_{t}$[GeV]
& $\Gamma^{(0)}_{nw}$[GeV] &
QCD[\%] & WBW[\%] & EW[\%] \\
\hline
 150.0& 0.809& 0.885& -8.47&-1.69 & 1.57\\
 160.0& 1.033& 1.130& -8.49&-1.60 & 1.62\\
 170.0& 1.287& 1.405& -8.49&-1.52 & 1.67\\
 180.0& 1.572& 1.714& -8.48&-1.45 & 1.70\\
 190.0& 1.890& 2.059& -8.47&-1.39 & 1.73\\
 200.0& 2.242& 2.440& -8.46&-1.33 & 1.76\\
\hline
\end{tabular*}
\end{table*}

\section{TOP QUARK DECAY}

\subsection{Total decay rate}
Top quark is the first heavy quark whose mass can
be measured to better than 1\% precision at a future $e^+e^-$
collider \cite{ExA}.
Therefore, measurements of its width can test not only
the Standard Model at the Born level, but
also the QCD radiative corrections
which are about -8.5\%.
This is in contrast to
$b$ and $c$ quarks, where uncertainties in
the masses and non-perturbative effects preclude
this possibility.\par
Being much heavier than the $W$ boson, the top quark decays
dominantly into $bW$ final states.
In the Standard Model
with three quark-lepton families, which is assumed throughout,
the modulus of the element $V_{tb}$
of the Cabibbo-Kobayashi-Maskawa quark mixing
\linebreak[4]
matrix
is close to 1 ( $|V_{tb}|$= 0.9985 to 0.9995 ) \cite{PDG}.
Neglecting small corrections due to the non-zero mass of $b$ quark
and the width of $W$ boson the total decay rate
can be written as follows:
\begin{equation}
\Gamma(t\to bW) =
{{{\rm G}_F} {m_t}^3\over 16\sqrt{2}\pi}
\left[ {\cal F}_0(y) - a_s {\cal F}_1(y)
\right]
\label{eq:totrate}
\end{equation}
where
\begin{equation}
a_s = {2\alpha_s\over 3\pi}
\end{equation}
\begin{equation}
y = \left( m_{\rm w}/m_t \right)^2
\end{equation}
\begin{equation}
{\cal F}_0(y) \;=\; 2(1-y)^2 (1+2y)
\label{eq:F0y0}
\end{equation}
and \cite{JK1}
\begin{eqnarray}
\lefteqn{
{\cal F}_1(y)  \,=\,
{\cal F}_0(y)\,
\left[\, {\textstyle {2\over3}}\pi^2
+4{\rm Li}_2(y)+2\ln y\ln(1-y)\,\right] }
\nonumber\\  &&
\hskip-5pt
      -\, (1-y)(5+9y- 6y^2)
      + 4y(1-y-2y^2)\ln y
\nonumber\\  &&
\hskip-5pt
      +\, 2(1-y)^2(5+4y)\ln(1-y)
\label{eq:F1y0}
\end{eqnarray}
The one loop electroweak corrections to
the total decay rate have been calculated
in the narrow width approximation \cite{DS}. They have
turned out to be positive and rather small (1-2\%).
A reason for this is that the contribution from the
Higgs--top quark Yukawa coupling remains very small for
realistic top quark masses \cite{IMT}.
The effect of the non-zero $W$
width  is comparable in size
to the electroweak correction but of the
opposite sign \cite{Topw}.
Formulae for the QCD corrected
total decay rate including non-zero $b$ quark mass
and $W$ boson width are given in \cite{JK1,Topw}.
In Table \ref{tab:totwid} we present
the results \cite{Topw} for the total decay rate
$\Gamma_t$ and its narrow width Born approximation
$\Gamma^{(0)}_{nw}$.
We give also the Standard Model contributions
to the width of the top quark
from the first order
QCD corrections, $W$ boson width (WBW)
and the electroweak (EW) corrections.
A number of intrinsic uncertainties remains.
It should be noted that the size of the electroweak corrections
is comparable to the uncertainties from the as yet uncalculated
${\cal O}({\alpha_s}^2)$ correction.
The present
uncertainty in $\alpha_s$ and the ignorance concerning
the second order QCD correction
limit the accuracy of the
prediction to about 1-2\%.
Experimental errors as well as theoretical uncertanties
in the determination of the top mass (c.f. Sect.\ref{threshold})
can also lead to effects of similar magnitude.

\subsection{Energy distributions}
\begin{figure}[htb]
\epsfxsize=7.5cm
\leavevmode
\epsffile[20 220 550 600]{mjfig2.ps}
\vskip-0.5cm
\caption{Distribution of $W$ energy for $m_t=$174 GeV without
(dashed line) and with (solid line) QCD corrections}
\label{fig-Wenergy}
\end{figure}
\subsubsection{Energy of $\bf W$}
The dominance of the two-body $t\to bW$ decay implies
that the energy of $W$ is strongly peaked.
In the top quark rest frame the two-body kinematics implies
\begin{equation}
E_{\rm w} = {m_t^2+m_{\rm w}^2-m_b^2\over 2m_t}
\end{equation}
The effects of non-zero $W$ width and QCD corrections have been
studied in \cite{CJKJ}.
In Fig.\ref{fig-Wenergy} the energy distribution
${\rm d}\Gamma/{\rm d}E_{\rm w}$ is
shown as the function of $E_{\rm w}$ for $m_t$=174 GeV.
The Dirac delta spike in $E_{\rm w}$ is smeared by
the effects of non-zero $W$ width, see dashed line.
QCD corrections result in further
distortion of the spectrum (solid line) because
hard gluon radiation leads to a long radiative tail and the virtual
corrections significantly reduce the height of the peak.

\subsubsection{Energy distributions of leptons}
The energy spectra of the leptons from semileptonic decays
of the top quark are sensitive to V-A form of $tbW$ vertex.
In first order QCD approximation
neglecting the masses of $b$ quark and
the leptons one derives the following normalized
energy distributions of the charged leptons and the neutrinos:
\begin{eqnarray}
\lefteqn{
{\rm A_l}\left(x_\ell\right) = {{\rm d}N\over{\rm d}x_\ell} =
12\,{ {\rm F}^+_0(x_\ell,y) - a_s {\rm F}^+_1(x_\ell,y)
\over
{\cal F}_0(y) - a_s {\cal F}_1(y)}  }
\nonumber\\ &&
\label{eq:enchar}\\
\lefteqn{
{\rm A}_\nu\left(x_\nu\right)= {{\rm d}N\over{\rm d}x_\nu} =
12\,{ {\rm F}^-_0(x_\nu,y) - a_s {\rm F}^-_1(x_\nu,y)
\over
{\cal F}_0(y) - a_s {\cal F}_1(y)}  }
\nonumber\\ &&
\label{eq:ennu}
\end{eqnarray}
for $y\le x_{\ell,\nu} \le 1$, where
$x_{\ell,\nu} = 2E_{\ell,\nu}/m_t$ denote the scaled energies in the
$t$ rest frame, the functions
${\cal F}_0(y)$ and ${\cal F}_1(y)$
have been defined in eqs.(\ref{eq:F0y0}) and (\ref{eq:F1y0}),
\begin{eqnarray}
{\rm F}^+_0(x,y) &=& x (1-x)
\label{eq:Fp0xy}\\
{\rm F}^-_0(x,y) &=& (x-y) (1-x+y)
\label{eq:Fm0xy}
\end{eqnarray}
and \cite{JK2,CJ}
\begin{eqnarray}
\lefteqn{
{\rm F}^+_1(x,y) =
   {\rm F}^+_0(x,y)\,\Phi_0 + x\Phi_1
   - ( 3 + 2x +y )\Phi_{2\diamond 3} }
\nonumber   \\   && \hskip-10pt
  + 5(1-x)\Phi_4 + ( - 2xy     + 9x     - 4x^2 - 2y - y^2)\Phi_5
\nonumber   \\
&&    \hskip-10pt   + y  (  4 - 4x - y + y/x )/2
\label{eq:Fp1xy}\\
\lefteqn{
{\rm F}^-_1(x,y) =
      {\rm F}^-_0(x,y)\,\Phi_0 + ( - 2xy  + x  + y  + y^2)\Phi_1  }
\nonumber\\
&&     \hskip-10pt
        + ( - 5  + 2x  - 3y )\Phi_{2\diamond 3}
        + (  5 + 4xy - 5x + 3y -
\nonumber\\
&&      \hskip-10pt
        5y^2   - 2y^2/x)\Phi_4
       + (  6xy + 9x  - 4x^2
     - 11y  - 2y^2
\nonumber\\  &&  \hskip-10pt
       + 2y^2/x )\Phi_5    + y  (  2 + 3x
        - 3y - 2y/x )/2
\label{eq:Fm1xy}
\end{eqnarray}
where
\begin{eqnarray}
&  \hskip-10pt
\Phi_0\ \ &\hskip-5pt
=\ {\textstyle {\pi^2\over 3}} + 2{\rm Li}_2(x)
 + 2{\rm Li}_2(y/x) + \ln^2\left({\textstyle {1-y/x\over 1-x}}\right)
\nonumber\\
&  \hskip-10pt
\Phi_1\ \ &\hskip-5pt
=\  {\textstyle {\pi^2\over 6}} + {\rm Li}_2(y)
           - {\rm Li}_2(x)   - {\rm Li}_2(y/x)
\nonumber\\
&  \hskip-10pt
\Phi_{2\diamond 3} &\hskip-5pt
=\ {\textstyle{1\over2}} (1-y)\ln(1-y)
\nonumber\\
&  \hskip-10pt
\Phi_4\ \ &\hskip-5pt
=\ {\textstyle{1\over2}}\ln(1-x)
\nonumber\\
&  \hskip-10pt
\Phi_5\ \ &\hskip-5pt
=\ {\textstyle{1\over2}}\ln(1-y/x)
\end{eqnarray}
\begin{figure*}[htb]
\epsfxsize=16.0cm
\leavevmode
\epsffile[60 300 490 525]{mjfig3.ps}
\vskip-1.5cm
\caption{Energy distributions a) ${\rm A_l}(x_\ell)$ of the charged
lepton and b) ${\rm A}_\nu(x_\nu)$ of the neutrino
for  the standard model V-A coupling ($\kappa^2=0$)
and an  admixture of V+A current
($\kappa^2=$0.1) for $y=$0.25 and $\alpha_s=$0.11. }
\label{fig-Ael}
\end{figure*}
\begin{table*}[hbt]
\caption{Moments of the energy distributions of the leptons
in  $t\to bW\to be^+\nu$
and the light quark jets in $t\to bW\to b\bar du$ decays.}
\label{tab:enmom}
\begin{tabular*}{\textwidth}{@{}c@{\extracolsep{\fill}}crrr}
\hline
  &  y         & $m_t$=150  &  $m_t$=175 &  $m_t$=200  \\
\hline   \\
$\left\langle x_\nu \right\rangle$  & ${1+4y+y^2\over2(1+2y)}$ &
   0.707     &   0.663   &     0.631    \\      \\
$\left\langle x_\ell \right\rangle$  & ${1+2y+3y^2\over2(1+2y)}$ &
   0.577     &   0.546   &     0.529    \\      \\
$\left\langle x_\nu- x_\ell \right\rangle$ & ${y(1-y)\over1+2y}$ &
   0.130     &   0.117   &     0.102    \\      \\
$\left\langle x_>- x_< \right\rangle$ & ${3(1+2y-3y^2)\over8(1+2y)}$ &
   0.317     &   0.340   &     0.353    \\
\\
\hline
\end{tabular*}
\end{table*}
Assuming the Standard Model
V-A structure of the charged current
the spectrum of the charged lepton vanishes at $x_\ell=1$
and the spectrum of the neutrino does not. The latter is also
significantly harder, see solid lines in Fig.\ref{fig-Ael}a-b.
The average (scaled) energy of the neutrino
$\langle x_\nu \rangle$
is  larger than
$\langle x_\ell \rangle$
of the charged lepton. For the sake of simplicity we neglect
QCD corrections and obtain the results shown in Table \ref{tab:enmom}.
It is interesting that for realistic range of $m_t$ the difference
$\langle x_\nu - x_\ell \rangle$
is significantly greater than zero.
This observation may be useful because a value close to zero
can be expected for the background $b+W$ events
of mass close to $m_t$ which
do not come from $t$ decays and are not correlated by the
dynamics.
We give also the V-A prediction for the average difference between
$x_>= 2E_>/m_t$ of the more energetic and
$x_<= 2E_</m_t$ of the less energetic lepton.
Let us remark that for hadronic decays, e.g.
$t\to bW\to b\bar d u$, up-type quarks play the role of
neutrinos and down-type antiquarks the role of charged leptons.
Thus,
$\langle x_> - x_< \rangle$
is also proportional to the average difference between
the energies of the more and less energetic jets for hadronic
decays of top quark. The values given in Table \ref{tab:enmom}
are much larger than the result of a crude estimation assuming
isotropic distribution of leptons in $W$ rest frame
and statistically independent $W$ and $b$ which implies
$\langle x_> - x_< \rangle_{bckg} =
{\textstyle {1\over4}}(1-y)$.
However, the former assumption
is not realistic in particular for $\bar pp$ colliders.

For V+A coupling the charged lepton and the neutrino energy
spectra would be interchanged in comparison to the V-A case.
In \cite{JK94} effects have been studied
of a small admixture of non-standard V+A current on
distributions of leptons. The $tbW$ vertex has been parametrized
as
\begin{eqnarray}
g_V\,\gamma^\mu\; +\; g_A\,\gamma^\mu\gamma_5
\label{eq:VA}
\\
g_V = ( 1+ \kappa)/\sqrt{1+\kappa^2}  \nonumber\\
g_A = ( -1+ \kappa)/\sqrt{1+\kappa^2} \nonumber
\end{eqnarray}
Hence $\kappa=0$ corresponds to pure V-A and $\kappa=\infty$
to V+A. In Fig. \ref{fig-Ael}a-b the lepton
spectra are plotted corresponding to $\kappa^2=0.1$, see
dashed lines. It can be seen that
the deviations from the results of the Standard Model (solid lines)
are rather small. They are larger for the polarization
dependent distributions which are discussed in the following.

\subsection{Chirality of $b$ jet}
A measurement of $b$ quark chirality in top decay offers
another opportunity to test the V-A form of $tbW$ vertex.
Chirality and helicity are nearly identical
for the highly relativistic $b$ quark which
originates from the decay,
so, a high degree of primordial polarization is expected.\\
It has been proposed long ago \cite{zerwas}
that distributions of charged leptons from semileptonic
decays of beautiful hadrons can be used in polarization
studies for $b$ quarks.
Some advantages of neutrino distributions have been
also pointed out \cite{CJKK,BR}.
Recently there has been considerable progress in the theory of
the inclusive semileptonic decays of heavy flavor hadrons.
It has been shown that in the leading order of an expansion
in inverse powers of heavy quark mass $1/m_Q$ the
spectra for hadrons coincide with those for the decays of
free heavy quarks \cite{CGG} and there are no $\Lambda_{QCD}/m_Q$
corrections to this result away from the energy
endpoint.
$\Lambda^2_{QCD}/m_Q^2$ corrections have been
calculated in \cite{Bigi,wise} for $B$ mesons and
in \cite{wise} for polarized $\Lambda_b$ baryons.
For some decays the results are similar to those of the well-known
$ACCMM$ model \cite{altar}.
Perturbative first order QCD corrections
to semileptonic decays of polarized $b$ quarks
are also known \cite{JK2,CJKK,CJ}. An old conflict
between  \cite{CCM} and \cite{JK2} can be considered
as solved in favor of the results of the latter work.\\
The theory is therefore in good shape. The real
drawback is that due to hadronization the net longitudinal
polarization of the decaying  $b$ quark is drastically
decreased. In particular those $b$ quarks become depolarized
which are bound in $B$ mesons
both produced directly and from $B^*\to B\gamma$ transitions.
Only those $b$'s (a few percent) which fragment directly
into $\Lambda_b$  baryons retain information on the original
polarization \cite{Bjo,CKPS}.
The signal is significantly reduced
and feasibility of polarization studies for $b$ jets is
problematic. The ongoing experimental studies
of $b$ polarization at LEP
should shed some light on this problem,
see \cite{AKV,Mele} for a review.
The reason is that $Z^0\to b\bar b$ decays
can be viewed as a source of highly polarized $b$ quarks
and the polarizations of $b$ jets in $Z^0$ and $t$ decays
are similar.
According to the Standard Model
the polarization of $b$ quarks from $Z^0$ decays
depends weakly on the production angle.
The degree of longitudinal polarization is fairly large,
amounting to $\langle P_b\rangle = -0.94$.
QCD corrections to this Born result are about 3\% \cite{KPT}.

\subsection{Polarized Top Quarks}

The analysis of polarized top quarks and their decays
has recently attracted considerable attention.
Studies at a linear electron-positron collider are particularly
clean for precision tests. However, also
hadronic \cite{DPRK,Kuehn4,KLY}
or $\gamma\gamma$ collisions \cite{FKK}
and subsequent spin analysis of top quarks might reveal new
information. These studies will result
in determination of the top
quark coupling to the $W$ and $Z$ bosons either
confirming the predictions
of the Standard Model
or providing clues for physics beyond.
The latter possibility is particularly intriguing for the top quark
because $m_t$
plays an exceptional role in the fermion mass spectrum.\par
A number of mechanisms have been suggested that will
lead to polarized top quarks.
In \cite{FKK}
this possibility has been discussed
for $\gamma\gamma$ collisions with circular polarized photons.
Related studies
may be performed in hadronic collisions which in this case, however,
are based on the correlation between $t$ and $\bar t$ decay products
\cite{Kuehn4,KRZ,KLY}. The most efficient and flexible
reactions producing
polarized top quarks are electron-positron collisions. A small
component of polarization transverse to the production plane
is induced by final state interactions which have been
calculated in perturbative QCD \cite{DPRK,KLY}.
The longitudinal polarization $P_L$
is large. Its dependence on the production angle, beam energy
and the top mass has been
discussed in \cite{KRZ,AS1}. $P_L$ varies
strongly with the production angle, e.g. between nearly $0.6$ for
$\cos\vartheta\,=\,-1$ and $-0.3$ for $\cos\vartheta\,=\,1$
at $\sqrt{s}\,=\,500$ GeV. Averaging over the production angle
leads therefore to a significant reduction of $P_L$ with typical
values of $\langle P_L\rangle$ around -0.2 \cite{KPT}.
QCD corrections
change $\langle P_L\rangle$ by a relative amount
of about 3\% \cite{KPT}. \\
All these reactions lead to sizable polarization and can be used
to obtain information on the production mechanism.
However, two drawbacks are evident:
production and decay are mixed in an intricate manner, and
furthermore the degree of polarization is relatively small
and depends on the production angle. Top quark production
with longitudinally polarized electron beams and close to
threshold provides one important exception: the restricted
phase space leads to an amplitude which is dominantly S-wave
such that the electron (and positron) spin is directly
transferred to the top quark. Close to threshold and with
longitudinally polarized electrons one may deal with
a highly polarized sample of top quarks {\em independent
of the production dynamics}. Thus one can study
$t$ decays under particularly convenient conditions:
large event rates, well identified restframe of the top quark,
and large degree of polarization.

\subsubsection{Angular distributions}
\begin{table*}[hbt]
\caption{Angular dependence of the distributions of $W$ bosons,
neutrinos and less energetic leptons in  $t\to bW\to be^+\nu$
or light quark jets in $t\to bW\to b\bar du$ decays.}
\label{tab:ang}
\begin{tabular*}{\textwidth}{@{}c@{\extracolsep{\fill}}crrr}
\hline
  &  y         & $m_t$=150  &  $m_t$=175 &  $m_t$=200  \\
\hline   \\
$h_\nu(y)$    & $1- {12y(1-y+y\ln y)\over(1-y)^2(1+2y)}$&
  -0.521      &  -0.311   &    -0.127    \\      \\
$h_{\rm w}(y)$& ${1-2y\over 1+2y}$   &
   0.275      &   0.410   &     0.515    \\      \\
$h_<(y)$& $1- {6y\{1-y-2y\ln[(1+y)/(2y)]\}\over(1-y)^2(1+2y)}$&
   0.464     &   0.509   &     0.559    \\
\\
\hline
\end{tabular*}
\end{table*}
In the rest frame of the decaying $t$ quark the angular distributions
of the decay products are sensitive to its polarization.
Let us define
the angle $\theta_{\rm w}$ between $W$ boson three-momentum and
the polarization three-vector $\vec s$. In the top quark rest
frame
$s = (0,\vec s\,)$ is the
polarization four-vector of the decaying top quark.
Note that $S= |\vec s\,| = 1$ corresponds to fully
polarized and $S=0$ to unpolarized top quarks. We define also
the angles $\theta_+$ and $\theta_0$ between $\vec s$ and the
directions of the charged lepton and the neutrino, respectively,
and $\theta_<$ for the less energetic lepton in semileptonic
or less energetic light quark in hadronic decays.
For the sake of simplicity let us
confine our discussion to Born approximation
and consider semileptonic $t\to bW\to b\ell^+\nu$
and hadronic $t\to bW\to b\bar d u$ decays.\\
The angular distribution of the charged lepton
is of the form
\begin{equation}
{ {\rm d}N\over{\rm d}\cos\theta_+} =
{1\over 2}\, \left[\, 1\,+
\,S\cos\theta_+ \right]
\label{eq:elec1}
\end{equation}
which follows, c.f. the following subsection, from
factorization of the angular-energy distribution
into an energy and an angular dependent part.
This factorization holds for arbitrary top mass below
and above the threshold for decays into real $W$ bosons
\cite{KS,JK2}. It is noteworthy that for $S$=1 the angular
dependence in (\ref{eq:elec1}) is maximal because
a larger coeffecient multiplying $\cos\theta_+$ would be
in conflict with positivity of the decay rate.
Thus the polarization analysing power of
the charged lepton angular distribution
is maximal and hence far superior to other
distributions discussed in the following. In particular
the angular distribution of the neutrino reads \cite{JK94}:
\begin{equation}
{ {\rm d}N\over{\rm d}\cos\theta_0} =
{1\over 2}\, \left[\, 1\,+
\,h_\nu(y)S\cos\theta_+ \right]
\label{eq:neut1}
\end{equation}
where  $h_\nu(y)$ is given in Table \ref{tab:ang}. The
distribution of the direction of $W$ can be easily obtained.
Only the amplitudes for the helicity states of $W$
$\lambda_{\rm w} = -1$ and $\lambda_{\rm w} = 0$
are allowed and their contributions to the decay
rate are in the ratio $\;2y\; :\; 1\;$ \cite{GilKau}.
The corresponding angular distributions are of the form
\begin{equation}
{{\rm d}N_{-1,0}\over{\rm d}\cos\theta_{\rm w}} = {1\over2}
\left(1\mp S \cos\theta_{\rm w} \right)
\end{equation}
After summation over the $W$ polarizations
the following angular dependence is obtained:
\begin{equation}
{{\rm d}N\over{\rm d}\cos\theta_{\rm w}} = {1\over2}
\left[1 + h_{\rm w}(y) S \cos\theta_{\rm w} \right]
\end{equation}
where  $h_{\rm w}(y)$ is also given in Table \ref{tab:ang}.
It is evident that the charged lepton
angular distribution is significantly
more sensitive towards the polarization of $t$ than the
angular distributions of $W$ and $\nu$.
The charged lepton is likely to be the less energetic
lepton because its energy spectrum is softer
than that of the neutrino.
For large values of $m_t$ the angular distribution
of the less energetic lepton
\begin{equation}
{{\rm d}N\over{\rm d}\cos\theta_<} = {1\over2}
\left[1 + h_<(y) S \cos\theta_< \right]
\end{equation}
is a more efficient analyser of top polarization than the angular
distribution of neutrinos. For $m_t$ in the range 150-200 GeV
it is also better than the direction of $W$,
c.f. Table \ref{tab:ang}.

\subsubsection{Angular-energy distributions}
For semileptonic decays
the normalized distributions of leptons
including first order QCD corrections
can be cast into the following form:
\begin{eqnarray}
{ {\rm d}N\over{\rm d}x_\ell\,{\rm d}\cos\theta_+} &=&
{1\over 2}\,\left[\,
{\rm A_l}(x_\ell)\, +\, S\cos\theta_+\,{\rm B_l}(x_\ell)\,
\right]
\nonumber\\
&& \\
{ {\rm d}N\over{\rm d}x_\nu\,{\rm d}\cos\theta_0} &=&
{1\over 2}\,\left[\,
{\rm A}_\nu(x_\nu)\, +\, S\cos\theta_0\,{\rm B}_\nu(x_\nu)\,\right]
\nonumber\\
\end{eqnarray}
where the functions ${\rm A_l}(x)$ and ${\rm A}_\nu(x)$ have
been defined in eqs. (\ref{eq:enchar}) and (\ref{eq:ennu}), and
\begin{eqnarray}
{\rm B_l}(x) &=& 12\,
 { {\rm J}_0^+(x,y)\, -\,
a_s\, {\rm J}_1^+(x,y)  \over
{\cal F}_0(y)\, -\, a_s\, {\cal F}_1(y)}
\label{eq:Bell}\\
{\rm B}_\nu(x) &=& 12\,
 { {\rm J}_0^-(x,y)\, -\,
a_s\, {\rm J}_1^-(x,y)  \over
{\cal F}_0(y)\, -\, a_s\, {\cal F}_1(y)}
\label{eq:Bnu}
\end{eqnarray}
where
\begin{eqnarray}
{\rm J}^+_0(x,y) &=& {\rm F}^+_0(x,y)
\label{eq:J0P}\\
{\rm J}^-_0(x,y) &=& (x-y) (1-x+y-2y/x)
\label{eq:J0M}
\end{eqnarray}
and \cite{CJK,CJ}
\begin{eqnarray}
\lefteqn{
{\rm J}^+_1(x,y) =
       {\rm J}^+_0(x,y)\,\Phi_0 - x\Phi_1
       +  ( 5 - 2x - y           }
\nonumber\\
&&     \hskip-0pt
        - 2y/x - 2/x )\Phi_{2\diamond 3}
       +  ( - 3 + x   + 2/x)\Phi_4
\nonumber\\
&&     \hskip-0pt
       +  ( - 2xy + 3x - 4x^2 + 6y - y^2 - 2y^2/x)\Phi_5
\nonumber\\
&&     \hskip-0pt
+  ( 2 - 2x^2 + 2y - 3y^2 - 2y/x + 3y^2/x )/2
\nonumber\\  &&
\label{eq:Jp1xy}\\
\lefteqn{
{\rm J}^-_1(x,y) =
      {\rm J}^-_0(x,y)\, \Phi_0
      + ( - 2xy - x  - 5y +  }
\nonumber\\
&&     \hskip-0pt
     y^2 - 2y^2/x )\Phi_1
      + ( 3 + 10x + y + 10y/x
\nonumber\\
&&     \hskip-0pt
      - 2/x )\Phi_{2\diamond 3}
      + ( - 3 + 12xy + x - 7y - y^2
\nonumber\\
&&     \hskip-0pt
    -  12y/x + 8y^2/x + 2/x )\Phi_4
    + ( 6xy - 9x
\nonumber\\
&&     \hskip-0pt
      - 4x^2   - y - 2y^2 + 10y^2/x)\Phi_5
        +  ( 2  - 5xy
\nonumber\\  &&
       - 2x^2  + 2y + 7y^2 - 2y/x - 2y^2/x )/2
\label{eq:Jm1xy}
\end{eqnarray}
\begin{figure*}[htb]
\epsfxsize=16.0cm
\leavevmode
\epsffile[60 300 490 525]{mjfig4.ps}
\vskip-1.5cm
\caption{Angular-energy distribution functions
in the Standard Model ($\kappa^2=0$)
and for the admixture of V+A current ($\kappa^2=$0.1):
a) ${\rm B_l}(x_\ell)$ for the charged
lepton and b) ${\rm B}_\nu(x_\nu)$ for the neutrino,
$y$=0.25 and $\alpha_s$=0.11.}
\label{fig:Bel}
\end{figure*}
Eq.(\ref{eq:J0P}) implies that in Born approximation
the double differential angular-energy distribution
of the charged lepton is the product of the energy
distribution and the angular distribution (\ref{eq:elec1}).
QCD corrections essentially do not spoil this factorization
\cite{CJK}. For the neutrino such factorization
does not hold, c.f. eqs.(\ref{eq:Fm0xy}) and (\ref{eq:J0M}).
After integration
over $x_\nu$ the angular dependence of the neutrino
distribution is much weaker than for the charged
lepton, c.f. eq.(\ref{eq:neut1}).\\
In Fig.\ref{fig:Bel} the functions
${\rm B_l}(x)$  and    ${\rm B}_\nu(x)$
are shown as solid lines for $y=$0.25 and $\alpha_s=$0.11 \cite{JK94}.
The effect of non-standard coupling defined in eq.(\ref{eq:VA})
is much stronger for the polarization dependent distribution
of neutrinos, see dashed lines in Fig.\ref{fig:Bel} corresponding
to $\kappa^2=0.1$ \cite{JK94}.
\begin{table*}[hbt]
\caption{The moments  ${\cal A}^{(\ell,\nu)}_k$ and
${\cal B}^{(\ell,\nu)}_k$ of the
angular-energy distributions for the charged leptons
and for the neutrinos
for $y=0.25$ and $\alpha_s= 0.11$ for $\kappa^2=0$
(upper entries)
and the ratios between the moments for
$\kappa^2=0.1$ and $\kappa^2=0$ (lower entries).}
\label{tab:angmome}
\begin{tabular*}{\textwidth}{@{}r@{\extracolsep{\fill}}rrrr}
\hline
   &          &          &          &     \\
$  k  $ &${\cal A}^{(\ell)}_k $ &${\cal B}^{(\ell)}_k$ &
${\cal A}^{(\nu)}_k$ & ${\cal B}^{(\nu)}_k$  \\
   &          &          &          &     \\
\hline
   -1   &    2.008  &    2.005  &   1.593  &  -0.707 \\
        &     .981  &     .940  &   1.023  &   1.166 \\
   &          &          &          &     \\
    0   &    1.000  &     .998  &   1.000  &  -0.452 \\
        &    1.000  &     .949  &   1.000  &   1.110 \\
   &          &          &          &     \\
    1   &     .559  &     .558  &    .683  &  -0.322 \\
        &    1.021  &     .960  &    .984  &   1.068 \\
   &          &          &          &     \\
    2   &     .345  &     .344  &    .500  &  -0.249 \\
        &    1.043  &     .973  &    .973  &   1.037 \\
   &          &          &          &     \\
    3   &     .230  &     .230  &    .385  &  -0.203 \\
        &    1.064  &     .989  &    .965  &   1.015 \\
\hline
\end{tabular*}
\end{table*}
In Table \ref{tab:angmome} the moments
\begin{eqnarray}
{\cal A}^{(\ell,\nu)}_k &=& \int_y^1{\rm d}y\, x^k\,
{\rm A}_{\ell,\nu}(x,y)     \nonumber\\
{\cal B}^{(\ell,\nu)}_k &=& \int_y^1{\rm d}y\, x^k\,
{\rm B}_{\ell,\nu}(x,y)
\label{eq:ABmom}
\end{eqnarray}
are given for integer $k$ between -1 and 3,
$y=0.25$ and $\alpha_s(m_t)= 0.11$ \cite{JK94} .
The upper entries in the table
denote the values of the moments for $\kappa^2=0$ and the
lower ones the ratios of the moments evaluated for
$\kappa^2=0.1$  to those  for $\kappa^2=0$.
It is evident that the
moments ${\cal B}_k^{(\nu)}$
which govern the angular dependence
of the neutrino spectrum
are particularly sensitive towards a V+A admixture. The effect
is most pronounced for the moment $k= -1$ which enhaces the lower
energy part of the spectrum and where the relative change amounts
to 17\% for $\kappa^2=0.1$~.
Thus, the angular-energy distribution of neutrinos
from the polarised top quark decay will allow for a particularly
sensitive test of the V-A structure of the charged current.
The effect of QCD correction can mimic a small admixture
of V+A interaction. Therefore, inclusion of the radiative
QCD correction to the decay distributions  is necessary for
a quantitative study.

\section{TOP QUARK PAIR PRODUCTION}

\label{threshold}
\begin{figure*}[htb]
\epsfxsize=8.0cm
\leavevmode
\epsffile[65 220 410 455]{mjfig5a.ps}
\epsfxsize=8.0cm
\leavevmode
\epsffile[65 220 410 455]{mjfig5b.ps}
\vskip-0.5cm
\caption{Cross section for $t\bar t$ production in units of
$\sigma_{point}$ including resonance and QCD enhancement
and initial state radiation.}
\label{fig:sigtot}
\end{figure*}
\subsection{$\bf e^+ e^- \to t \bar{t}$}

It is evident that the bulk of top studies
at an $e^+ e^-$ collider will rely on quarks produced
in $e^+ e^-$ annihilation
through the virtual $\gamma$ and Z, with a production cross
section of the
order of $\sigma _{point}$\footnote{This subsection
is a shortened version of a comprehensive review given
in \protect\cite{ThA}.}.
For quarks tagged at an angle $\vartheta $, the
differential cross section in Born approximation
is a binomial in $\cos \vartheta $
\begin{eqnarray}
\frac{d\sigma}{d \cos\vartheta}
&=& \frac{3}{8} \left( 1 + \cos^2 \vartheta \right) \sigma _U
+ \frac{3}{4} \sin^2 \vartheta \sigma _L
\nonumber\\  &&
+ \frac{3}{4} \cos \vartheta \sigma _F
\label{eq:eetott1}
\end{eqnarray}
$U$ and $L$ denote the contributions of unpolarized
and longitudinally polarized
gauge bosons along the $\vartheta$ axis,
and $F$ denotes the difference
between right and left polarizations.
The total cross section is the sum
of $U$ and $L$:
\begin{equation}
\sigma = \sigma _U + \sigma _L
\label{eq:eetott2}
\end{equation}
In Born approximation
the coefficients $\sigma^i$ can be expressed in terms of
the cross sections for the
massless case
\begin{eqnarray}
\sigma ^U_B & = & \beta \sigma^{VV}
+ \beta ^3 \sigma ^{AA}
\\
\sigma ^L_B & = & \frac{1}{2} \left( 1 - \beta ^2 \right)
\beta \sigma^{VV}
\\
\sigma ^F_B & = & \beta ^2 \sigma^{VA}
\label{eq:eetott4}
\end{eqnarray}
where
\begin{equation}
\beta=\sqrt{1- 4m_t^2/s}
\end{equation}
\begin{eqnarray}
\lefteqn{
\sigma ^{VV}  =  \frac{4 \pi \alpha^2 (s)
e^2_e e^2_Q}{s}
}
\nonumber \\
& & + \frac{G_F \alpha(s)}{\sqrt{2}} e_e e_Q
(\upsilon _e + \rho a_e )
\upsilon _Q \frac{m^2_Z \left( s-m^2_Z \right)}
{D\left(s,m_Z,\Gamma_Z\right)}
\nonumber \\
& & + \frac{G^2_F}{32 \pi} \left( \upsilon ^2_e
+ a^2_e + 2 \rho \upsilon _e a_e \right) \upsilon ^2_Q \frac{m^4_Z s}
{D\left(s,m_Z,\Gamma_Z\right)}
\nonumber \\
\lefteqn{
\sigma ^{AA}  =  \frac{G^2_F}{32 \pi}
\left( \upsilon ^2_e + a^2_e + 2 \rho \upsilon _e a_e \right)
a^2_Q \frac{m^4_Z s}
{D\left(s,m_Z,\Gamma_Z\right)}
}
\nonumber \\
\lefteqn{
\sigma ^{VA}  =  \frac{G_F \alpha (s)}{\sqrt{2}} e_e
(\rho \upsilon _e + a_e )e_Q a_Q \frac{m^2_Z
\left( s-m^2_Z \right)}
{D\left(s,m_Z,\Gamma_Z\right)}
}
\nonumber\\
& & + \frac{G^2_F}{16 \pi} \left[ 2 \upsilon _e a_e
+ \rho (\upsilon ^2_e + a^2_e)\right]
\frac{
\upsilon_Q a_Q \;
m^4_Z s}
{D\left(s,m_Z,\Gamma_Z\right)}
\nonumber\\  &&
\label{eq:eetott5}
\end{eqnarray}
and
\begin{equation}
{D\left(s,m_Z,\Gamma_Z\right)}  =
{\left( s-m^2_Z\right)^2
+ \left(s\Gamma _Z/{m_Z} \right)^2}
\nonumber
\end{equation}
The fermion couplings are given by
\begin{eqnarray}
\upsilon _F = 2I^f_3 - 4e_f \sin ^2 \theta _w \quad ,
\qquad a_f = 2I^f_3
\label{eq:eetott6}
\end{eqnarray}
and the possibility of longitudinal electron polarization
($\rho = -1;+1;0$ for
righthanded; lefthanded; unpolarized electrons) has been included. \\
QCD corrections to this formula are available
for arbitrary $m^2 / s$ up
to first order in $\alpha _s$:
\begin{eqnarray}
\sigma & = & \frac{(3 - \beta ^2)}{2}
\beta \sigma^{VV}\,R_V
+ \beta ^3 \sigma ^{AA}\,R_A
\label{eq:eetott7}
\end{eqnarray}
The exact result \cite{exac}
for the QCD enhancement factors can be
well approximated by \cite{approx}
\begin{eqnarray}
R_V & = & 1 + \frac{4\alpha_s}{3}
\left[ \frac{\pi}{2\beta} - \frac{3+\beta}{4}
\left( \frac{\pi}{2} - \frac{3}{4\pi} \right) \right]
\label{eq:eetott10b}
\end{eqnarray}
and
\begin{eqnarray}
\lefteqn{ R_A  =  }
\nonumber\\  &&    \hskip-20pt
 1 + \frac{4\alpha_s}{3}
\left[ \frac{\pi}{2\beta} -
\left(\frac{19}{10} - \frac{22}{5}\beta +\frac{7}{2}\beta^2\right)
\left( \frac{\pi}{2} - \frac{3}{4\pi} \right) \right]
\nonumber\\
\label{eq:eetott10c}
\end{eqnarray}
For small $\beta$ these factors develop the familiar Couloumb
enhancement $\sim{{2\over3}\pi\alpha_s/\beta}$ compensating the
phase space suppression
$\sim \beta$.  This leads to a nonvanishing cross
section which smoothly joins the threshold region.\\
Initial state radiation has an important influence on the magnitude
of the cross section
which is significantly suppressed in
the threshold region. The correction factor increases rapidly
with energy, but stays below 0.9 in
the full range under consideration.
In Fig.\ref{fig:sigtot} the cross section for
$e^+e^-\to t\bar t$ is shown as the function of the center-of-mass
energy \cite{ThA}.

\subsection{Pair production near energy threshold}

It follows from Table \ref{tab:totwid} that the
top quark is a short--lived particle. Its width
$\Gamma_t$ is in the range 1--2 GeV and fairly exceeds
the tiny hyperfine splitting for open top hadrons,
the hadronization scale of about 200 MeV,
and even the energy splitting between $1S$ and $2S$ $t\bar t$
resonances. On one side this is an advantage because
long-distance phenomena related to confinement are
apparently less important for top quarks \cite{BDKKZ}.
In particular depolarization due to hadronization is
practically absent.
On the other side the amount of information available from
the threshold region is significantly reduced. Toponium
resonances including $1S$ overlap each other.
As a consequence the cross section
for $t\bar t$ pair production
near energy threshold has a rather simple and smooth shape.
Nevertheless, as first pointed out by Fadin and Khoze \cite{FK}
the excitation curve $\sigma(e^+e^-\rightarrow t\bar t\,)$
allows a precise determination of $m_t$ as well as of other
physical quantities such as $\Gamma_t$ and the strong coupling
constant. The results of \cite{FK}
for the Coulomb chromostatic potential
have been derived analytically
in the non-relativistic approximation.
Strassler and Peskin \cite{SP} have obtained similar results
using numerical approach and a more realistic
QCD potential.
The idea \cite{FK,SP} to use Green function
instead of summing over overlapping resonances has been also
applied in calculations of differential cross sections
\cite{Sumino1,JKT,JT,Martinez,Sumino2}.
Independent numerical approaches have been devoleped for
solving Schr\"odinger equation in the position space \cite{Sumino1}
and Lippmann-Schwinger equation in the momentum space \cite{JKT,JT}.
The results of these two methods agree very well \cite{Martinez}.
One of the most important future applications
will be in determination of $m_t$ and $\alpha_s$.
\begin{figure*}[htb]
\epsfxsize=8.0cm
\leavevmode
\epsffile[10 205 585 640]{mjfig6a.ps}
\epsfxsize=8.0cm
\leavevmode
\epsffile[10 205 585 640]{mjfig6b.ps}
\vskip-0.5cm
\caption{Distribution
$|p{\cal G}(\vec p,E)|^2$%
of the momentum $p=|\vec p|$ of the top quark %
in $t\bar t$ system %
for $E$=-2.9 GeV ($1S$ peak), $E$=0 and $E$=2 GeV
a) $m_t$= 150 and b) 180 GeV.}
\label{fig:Green}
\end{figure*}
It has been argued
\cite{Sumino1} and demonstrated \cite{Martinez,Sumino2} that
the combined measurements of the total
and the differential cross sections in
$e^+ e^- \to t\bar t$
offer a very promising method for a simultaneous
determination of $m_t$ and $\alpha_s(m_t)$.
The estimated errors of such a determination are \cite{Martinez}:
$\Delta m_t$=300 MeV and $\Delta\alpha_s$=0.006 for $m_t$=150 GeV and
$\Delta m_t$=520 MeV and $\Delta\alpha_s$=0.009 for $m_t$=180 GeV.
The errors are correlated, so for
$\alpha_s$ fixed by other measurements a much better
determination of $m_t$ can be obtained. In the following
the method of \cite{JKT,JT} is presented. Some theoretical
uncertainties related to the momentum dependent width of $t\bar t$
system, long-distance part of the QCD potential
and the definition of the pole top quark mass are discussed.
Effects of Higgs exchange are also considered.

\subsubsection{Green function}
Let us describe briefly the Green function method
for $e^+e^-\rightarrow t\bar t$ annihilation
near the energy threshold
and the numerical solution of the Lippmann--Schwinger equation;
see \cite{JKT,JT} for details.
We simplify the original calculations and
neglect the $Z^0$ contribution and the transverse
gluon correction to the production vertex.
The differential cross section
for the top quark pair production
in electron-positron annihilation reads:
\begin{equation}
{{\rm d}\sigma\over {\rm d}^3p} \left(\vec p,E\right) =
{3\alpha^2\,Q_t^2\over\pi\,s\, m_t^2}
\Gamma(p,E)\left|{\cal G}(\vec{p},E)\right|^2\
\label{eq:III1}
\end{equation}
where $\Gamma(p,E) = \Gamma_{t \bar t}/2$~.
The Green function ${\cal G}(\vec{p},E)$ is the solution of
the non-relativistic Lippmann-Schwinger equation
\begin{eqnarray}
\lefteqn{
{\cal G}(\vec{p},E) =
{\cal G}_0(\vec{p},E) + }
\nonumber\\   && \hskip-15pt
{\cal G}_0(\vec{p},E)
\int {{\rm d}^3q\over(2\pi)^3}
\tilde V(\vec{p}-\vec{q}\,)
{\cal G}(\vec{q},E)
\label{eq:III2}
\end{eqnarray}
where
\begin{equation}
E = \sqrt{s} - 2m_t
\end{equation}
denotes the total energy of the $t\bar t$ system and
$\vec p$ the momentum of $t$ quark.
$\tilde V(\vec{p}\,)$ is the potential in momentum space
and  $\Gamma(p,E)=\Gamma_{t\bar t}/2$,
where  $\Gamma_{t\bar t}$ denotes the width of the
the $t\bar t$ system.
The free Hamiltonian that is used to define the Green function
${\cal G}_0$
includes the momentum dependent width:
\begin{equation}
{\cal G}_0(\vec{p},E)
= {1\over E- {p^2\over m_t}+ {\rm i}\Gamma(p,E)}
\label{eq:III8}
\end{equation}
Near the energy threshold one can neglect all but $S$ partial
waves and solve numerically the corresponding one--dimensional
integral equation. The spherically symmetric
solution fulfills the unitarity condition
\cite{Sumino1}
\begin{eqnarray}
\lefteqn{
\int \frac{{\rm d}^3p}{(2\pi)^3}\,
\Gamma(p,E)\,\left|{\cal G}(p,E)\right|^2 =}
\nonumber\\  &&  \hskip15pt
- {\rm Im} G(\vec x=0,\vec x\,^{\prime}= 0,E)
\label{eq:III9}
\end{eqnarray}
which for the constant decay rate reduces to
the formula for the total cross section  derived
in \cite{FK,SP}. Eq.(\ref{eq:III9}) can be also considered
as a non-trivial cross check of the numerical approach.\\
In Fig.\ref{fig:Green} the momentum distributions
of $t$ quarks in $t\bar t$ systems of energy $E$ corresponding
to $1S$ peak , $E$=0 and $E$=2 GeV
are shown
as solid, dotted and
dashed lines respectively.
The shift towards larger momenta with increasing $E$ and
the narrowing of the distributions is clearly visible.
It has been shown \cite{Martinez,Sumino2} that the position
of the maximum is not sensitive to the initial state radiation,
so it can be used to measure $E$ and $m_t$. On the other hand
the rapid increase of the cross section
$\sigma(e^+e^-\to t\bar t\,)$ is related
to the location of $1S$ resonance which depends on both
$m_t$ and $\alpha_s$. Thus, the combined measurement of
both the total and differential cross sections leads
to a simultaneous determination of these parameters.
The results shown in Fig.\ref{fig:Green} have been obtained
in \cite{JKT} assuming constant width of $t\bar t$ system
$\Gamma(p,E)=\Gamma_t$. Let us discuss now theoretical
problems which appear when this assumption is lifted.

\subsubsection{Width of $t \bar t$ system.}
The width of the $t\bar t$ system depends on
the momentum of $t$ quark because both the
matrix element and the phase space available for the decay products
depend on it. When produced near energy threshold
$t$ and $\bar t$ cannot be considered
as free particles. The binding energy and the
kinetic energy  of internal motion
tend to reduce the available
phase space for the decay. Although the effect is only
${\cal O}({\alpha_s}^2)$ the suppression
can be significant \cite{JK}.
Thus, in a high precision calculation
one has to consider the width $\Gamma_{t-\bar t}(p)$ as
a non-trivial function of the momentum $p$.
\begin{figure}[htb]
\epsfxsize=7.0cm
\leavevmode
\epsffile[30 220 535 590]{mjfig7.ps}
\vskip-0.5cm
\caption{Comparison of the annihilation cross section
$\sigma(e^+e^-\to t\bar t)$ evaluated for the constant width
$\Gamma_t$ (dashed line) and the momentum dependent widths
including time dilatation (solid line) and phase space
suppression (dotted line).}
\label{fig:mdw}
\end{figure}
The phase space effect tends to reduce the decay rate
of bound top quarks relative to free ones \cite{JK,Sumino1} and the
effect is enhanced because for short--lived
particles the momentum distribution is broad.
However, for the same reason the decays
take place at short relative distances, where the wave functions
of $b$ and $\bar b$ quarks originating from the decays are
distorted ({\em enhanced}) by Coulomb attraction. Therefore, when
calculating the amplitude of $t\rightarrow bW$ transition,
one should use Coulomb wave functions rather than plane waves
for $b$ quarks. This effect clearly increases the rate. A third
factor is due to time dilatation: a top quark moving with velocity
$v$ lives longer in the center--of--mass frame.
While phase space reduction and time dilatation can be
implemented in a straightforward way Coulomb enhancement
cannot be easily taken into account. In principle
one has to replace the plane wave functions for $b$ quarks by
relativistic Coulomb functions when evaluating the amplitude
for the $t\rightarrow bW$ transition.
In Fig.\ref{fig:mdw} \cite{JT} predictions for
the annihilation cross section
$\sigma(e^+e^-\to t\bar t\,)$
are shown corresponding to different asumptions about
$\Gamma_{t\bar t}$:
solid line has been obtained including only time dilatation,
dotted line including only the phase space suppression
and the dashed line corresponds to the constant width.
The effect of the phase space suppression is non-negligible
whereas time dilatation produces rather small effect.\\
It has been conjectured \cite{JKT,JT,Pilk} that,
in close analogy to what happens in the case of negative muons
bound in nuclei \cite{Huff},
for chromostatic attraction in $t\bar t$ systems the phase
space suppression and the Coulomb  enhancement
cancel each other and only the suppression
due to time dilatation should be included.
Recently this conjecture has been proven \cite{Kummer}.
Moreover, it has been explicitly demonstrated that
unphysical gauge dependence of the momentum dependent
width \cite{Sumino1} also disappeares in the final
result \cite{Kummer}. The distributions
of the top quark momentum, see  Fig.\ref{fig:Green},
are also only weakly affected.
It should be stressed, however,
that the experimentally accessible
three-momentum distributions of $Wb$
systems from top quark decays are not identical to the
distributions of the top quark momentum.
Calculable corrections arise due to final state
rescattering \cite{Sumino2}.

\subsubsection{QCD potential}
\begin{figure}[htb]
\epsfxsize=7.0cm
\leavevmode
\epsffile[65 175 525 655]{mjfig8.ps}
\vskip-0.5cm
\caption{$\alpha_{eff}(q)$ for different values of
$\alpha_s(m_{\rm z})$:
solid -- 0.12, dashed -- 0.11, dash-dotted -- 0.13 and
dotted -- 0.10 and 0.14.}
\label{fig:alefq}
\end{figure}
\begin{figure}[htb]
\epsfxsize=7.5cm
\leavevmode
\epsffile[70 80 500 420]{mjfig9.ps}
\vskip-0.5cm
\caption{QCD potential $V_{JKT}(r)$ for different values of
$\alpha_s(m_{\rm z})$:
solid -- 0.12, dashed -- 0.11, dash-dotted -- 0.13 and
dotted -- 0.10 and 0.14.}
\label{fig:VJKTR}
\end{figure}
Some difficulties appear also for the QCD potential.
In order to illustrate these problems
let us consider
the following potential \cite{JT}
defined in momentum space ($p=|\vec p\,|$):
\begin{equation}
\tilde V_{JKT}(p) = - \frac{16\pi}{3} \frac{\alpha_{eff}(p)}{p^2} +
V_0\,\delta(p)
\label{eq:VJKTp}
\end{equation}
The function
\begin{eqnarray}
\alpha_{eff}(p)=\left\{
      \begin{array}{ll}
       \alpha_{pert}(p) & \qquad {\rm if}\ p>p_1 \\
       \alpha_R(p) & \qquad {\rm if}\ p<p_2 \\
      \end{array} \right.
\label{eq:alphaeff}
\end{eqnarray}
is shown in Fig.\ref{fig:alefq}.
At large momenta, i.e. for $p>p_1$=5 GeV
$V_{JKT}(p)$ is equal to
the perturbative two-loop QCD potential \cite{potential}
for $n_f=5$ quark flavours and $\alpha_s(m_{\rm z})$= 0.10--0.14.
At intermediate and small momenta,
i.e. for $p<p_2$=2 GeV a Richardson--like
phenomenological potential \cite{Richardson} is employed,
and a linear interpolation formula is used in between.
While the form of the potential in momentum space
is fixed at large and intermediate momenta
by perturbative QCD and phenomenology of $b\bar b$ and $c\bar c$
quarkonia, it is not known in the infrared (confinement) region.
Thus $V_0$ in eq.(\ref{eq:VJKTp})
is to some extent an arbitrary parameter. In \cite{JK} $V_0$
has been
fixed by imposing an evidently arbitrary condition
on the potential in the position space
\begin{equation}
V_{JKT}(r=1 {\rm GeV}^{-1}) =  - 1/4 {\rm GeV}
\label{condition}
\end{equation}
Then after Fourier transformation the potential $V_{JKT}(r)$
is obtained, see Fig.\ref{fig:VJKTR},
which describes well the $b\bar b$ and $c\bar c$ quarkonia.
A different choice of $V_0$ results in a redefinition
of $b$ and $c$ quark masses which
in the framework of the non-relativistic
approach are defined only with limited ($\sim$ 300 MeV)
precision.
In particular one can impose conditions relating
$V_0$ and $\alpha_s(m_{\rm z})$ which lead to different
energy dependence of the momentum distributions
in $t\bar t$ systems.
At this point a brief discussion is in order on the relation
between the top masses determined on the basis of different
potentials \cite{JKT1}.  The perturbative two-loop QCD
potential in momentum space is fixed
unambiguously for sufficiently large $Q^2$.
In order to calculate the potential in
coordinate space the small $Q^2$ behavior has to be specified in
an ad hoc manner.  Different assumptions will lead to the same
short distance behavior.  The potentials will, however, differ
with respect to their long distance behavior.  In \cite{FK} it
has been argued convincingly that the long distance tail is cut off
by the large top width.  However, an additive constant in coordinate
space can be induced by the small momentum part of $\tilde V(p)$.
This additional term leads to a shift in the $t\bar t$ threshold,
which in turn can be reabsorbed by a corresponding shift in $m_t$.
The different assumptions on the long distance behavior are reflected
in differences between
the predictions of \cite{SP,Sumino1,JT}
for the precise location of the $t\bar t$
threshold for identical values of $\alpha_s$
and $m_t$ as well as in differences
in the $\alpha_s$ dependence of the momentum distributions for fixed
$m_t$ and energy, see also \cite{Martinez}.
All these differences can be attributed to the
freedom in the additive constant discussed before.  The same additive
constant appears in $b\bar b$ spectroscopy, such that the mass
difference between top and bottom is independent from these
considerations. Recently problems with the definion
of $b$ quark mass due to long distance effects (infrared renormalons)
have been noted in \cite{Braun}.\\
\begin{figure}[htb]
\epsfxsize=7.5cm
\leavevmode
\epsffile[20 40 340 300]{mjfig10.ps}
\vskip-0.5cm
\caption{Comparison of $t\bar t$ cross section as the function
of energy relative to the position of $1S$ peak for different
potentials \protect\cite{HiSu}.}
\label{fig:HiSu}
\end{figure}
A similar study has been performed in \cite{HiSu}.
In Fig.\ref{fig:HiSu}
$\sigma(e^+e^-\to t\bar t\,)$ is shown as the function of
energy relative to the location of the $1S$ peak
$\Delta E = \sqrt{s} - E_{1s}$. The solid lines have been
obtained for the potential of \cite{Sumino1} and the dashed ones
for the potential of \cite{SP}. These two potentials are equal
at short distances but different prescriptions are used
at intermediate and large distances. It can be seen that the height
of the $1S$ peak is affected. The ambiguity may be fixed by
performing a simultaneous fit to the differential momentum
distribution, angular distributions (e.g. the forward-backward
asymmetries) and/or by fixing $\alpha_s$ from independent
measurements. It is clear that this problem deserves a careful
further study.

\subsubsection{Higgs effects in the threshold region}
Effects of Higgs exchange may be non-negligible in the
threshold region\cite{SP}
in particular if the Higgs mass is relatively low as predicted
by supersymetric models.
The dominant effect from a light Higgs boson can be
described by an instantaneous Yukawa potential
\begin{equation}
V_{\it Yuk}(r)=-\kappa\frac{e^{-m_h r}}{r}
\end{equation}
with
\begin{equation}
\kappa=\sqrt 2 G_F m_t^2/4\pi
\end{equation}
The vertex correction from Higgs exchange
can be included in the potential
calculation by adding $V_{\it Yuk}$ to the QCD potential:
\begin{equation}\label{subtr}
          V_{eff} = V_{QCD} + V_{\it Yuk}
\end{equation}
Then, $V_{eff}$ can be used instead of $V_{QCD}$ in numerical
calculations of the total \cite{SP} and differential
\cite{Sumino1,JKT} cross sections for $e^+e^-$ annihilation
near $t\bar t$ threshold.
The dominant contribution to the remainder
can be included as an overall correction factor
to the amplitude \cite{JKHi}:
\begin{equation} \label{fVhat}
1+\frac{\kappa}{\pi} \left[
f_{thr}(m_h^2/m_t^2) - \pi \frac{m_t}{m_h}
\right]
\end{equation}
where \cite{GKKS,guth,Fadin}
\begin{eqnarray}\label{fVthr}
\lefteqn{
f_{thr}(r)=-\frac{1}{12}\left[-12+4r+(-12+9r-2r^2)\ln r
\right.}
\nonumber\\ &&
\left.     +   \frac{2}{r}(-6+5r-2r^2)l_4(r) \right]
\end{eqnarray}
\begin{eqnarray}
\lefteqn{
l_4(r)=
\left\{
\begin{array}{ll}
\sqrt{r(4-r)}\,\arccos(\sqrt r /2) & \hskip-2pt
\mbox{\quad if $r\le4$}\\
 -\sqrt{r(r-4)}\,\frac{1}{2}\ln\frac{1+\sqrt{1-4/r}}{1-\sqrt{1-4/r}}
        & \hskip-2pt \mbox{\quad if $r>4$}
\end{array}
\right. }
\nonumber\\
&&
\end{eqnarray}
This factor
is energy independent and can be used below as well as
above the threshold.
\vskip0.5cm

\noindent{\bf Acknowledgements}\\

\noindent
I would like to gratefully acknowledge the collaboration
with
A.~Czarnecki, C.~J\"unger,
J.H.~K\"uhn and T. Teubner
on many aspects of top quark
physics covered in this review.
I thank I.~Bigi, E.~Boos, A.~Buras, K.~Hagiwara,
P.~Igo-Kemenes, V.~Khoze,
M.~Martinez, R.~Miquel, M.~Peskin, H.~Pilkuhn,
M.-L.~Stong, Y.~Sumino, M.~Veltman, K.~Zalewski and P.~Zerwas
for useful comments and discussions.

\end{document}